\documentclass[pra,twocolumn,notitlepage,superscriptaddress,longbibliography]{revtex4-2}
\usepackage[unicode=true,colorlinks=true,citecolor=blue,urlcolor=blue]{hyperref}
\usepackage{bm}
\usepackage{epsfig}
\usepackage{amsmath}
\usepackage{multirow}

\usepackage[normalem]{ulem}

\renewcommand {\phi}{{\varphi}}
\newcommand {\rmi}{{\rm i}}

\newcommand {\e}{{\rm e}}
\newcommand {\eps}{\varepsilon}

\begin{document}
\title{Van der Waals interactions in linear chains with long-range correlated disorder }

\author{Yuval Mualem}

\address{Department of Physics of Complex Systems, Weizmann Institute of Science, Rehovot 7610001, Israel and Department of Chemistry, Institute of Nanotechnology \& Advanced Materials, Bar-Ilan University, 5290002 Ramat-Gan, Israel}
\email{yuvalm225@gmail.com}

\author{Alexander N. Poddubny}

\email{poddubny@weizmann.ac.il}

\address{Department of Physics of Complex Systems, Weizmann Institute of Science, Rehovot 7610001, Israel}

%

\begin{abstract}
We present a theoretical study of van der Waals interaction forces in disordered linear molecule chains. We demonstrate that the 
interaction energy strongly and nonmonotonously depends on 
the disorder correlation length. Semianalytical expressions for the interaction energy are obtained. 
\end{abstract}
\maketitle 

\section{Introduction}\label{sec:intro}
Casimir and van der Waals interactions are essentially the electromagnetic interactions between the fluctuating dipole moments~\cite{French_2010,Kats_2015,leonhardt2015forces,Woods_2016,Fiedler2023}.  It is hard to overestimate their ubiquity in various systems. To give a few examples, these interactions underpin the surface tension effects in liquids, the formation of layered van der Waals materials actively studied in condensed matter physics~\cite{Geim2013}, and  provide the nonlinearities in quantum optics~\cite{Lukin2001,Drori2023}. More recently, it has been proposed that the fluctuating Casimir-type interactions may also be used as an alternative hypothesis to the dark matter, explaining the expansion of the Universe~\cite{leonhardt2020}. 

The theoretical approaches to calculating the impact of these forces in various systems are also well established. These range from the atomistic approaches, where the discrete structure of the interacting objects is rigorously taken into account either in a classical or in a full quantum mechanical way~\cite{Woods_2016}, to the classical approach of Lifshitz, where the problem is reduced to electrodynamics of continuous media~\cite{Lifshitz1956,landau9}. The classical electrodynamics approach is very efficient to provide the forces {\it between} the two separate objects. It does not require any microscopic knowledge of the atomic structure, and the system is entirely characterized by the dielectric permittivity $\eps(\bm r,\omega)$, depending on the position $\bm r$ and the frequency $\omega$. {The problem of calculating the force between the two objects embedded in the medium  is still  not trivial.  
This is somewhat similar to another well-known  problem of local field corrections, arising when calculating the spontaneous emission of an atom embedded inside a medium~\cite{Scheel2008}. The calculated force between two objects depends on the definition of the stress tensor in a medium~\cite{Burger2018,Burger2020}, which has been a subject of intense debates~\cite{Obukhov2003,Richter2008,Toptygin2016}. A recent comparison with experimental data in Ref.~\cite{Burger2020}  favors the standard continuous-medium Lifshitz theory. However,
the Lifshitz theory encounters yet another problem, inside an {\it inhomogeneous} medium:  the Maxwell stress tensor diverges~\cite{simpson2014surprises}.
}
In the traditional approach of Dzyaloshinskii and  Pitaevskii~\cite{landau9}, it is assumed that the medium is piecewise-homogeneous, $\eps(\bm r)=\rm const$, and the divergent term can be subtracted. This subtraction procedure is not well defined for a non-homogeneous medium, and a more sophisticated regularization is required. 
Various regularization procedures to the Casimir and van del Waals interactions have been proposed~\cite{Philbin_2010,Griniasty_2017}, see also references in Ref.~\cite{Griniasty_2017}
A rigorous verification of the phenomenological regularization has recently been performed in Ref.~\cite{Leonhardt2023}.
The authors considered a toy discrete model of point dipoles along a line, mimicking an actual discrete medium.
{We note that such type of toy dipole models also apply to van der Waals interactions in various more realistic systems. Examples of related problems  include  light harvesting in organic molecules~\cite{Zech_2014},
 topological spin phases of trapped exciton arrays \cite{Poddubny2019}, quantum physics  of spin networks
\cite{Christandl2005} and electronic structure of  defect chains in solids~\cite{Ghassemizadeh2022,Minke2024}.  
In Ref.~\cite{Leonhardt2023}, the dipole model was used to rigorously justify  the phenomenological regularizations of the Casimir and van der Waals forces, previously introduced in Ref.~~\cite{Griniasty_2017}.}

\begin{figure}[b]
\centering\includegraphics[width=\linewidth]{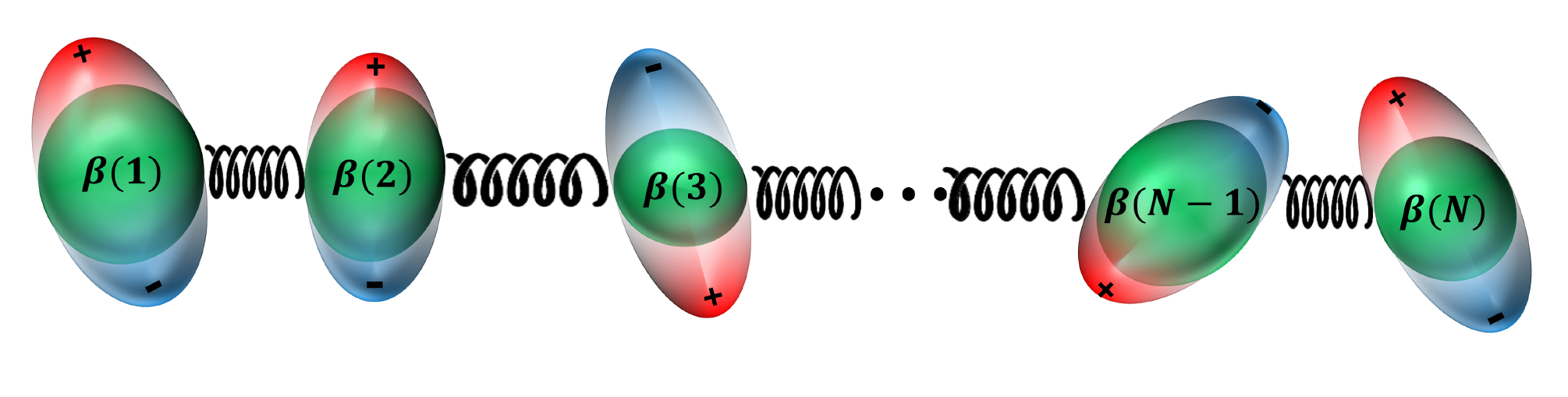}
\caption{Schematic illustration of   $N$ particles with different strengths of polarizability $\beta$.
} \label{fig:schematics}
\end{figure}

These recent studies indicate that even the problem of van der Waals and Casimir interactions in the one-dimensional chain is not trivial and can provide interesting insights. Here, we consider these interactions in the disordered one-dimensional array of particles with fluctuating dipole susceptibilities, illustrated in Fig.~\ref{fig:schematics}. We calculate the total Casimir energy in such a system and find that it depends in a non-monotonous way on the correlation length of the disorder. An analytical approximation in the limit of smooth disorder is proposed, and it is found to satisfactorily describe the numerical results.

The rest of the manuscript is organized as follows. Section~\ref{sec:model} presents our theoretical model. We start by presenting the basic equations for the Casimir interaction energy in the homogeneous array in Sec.~\ref{sec:model:hom} and then discuss the role of array inhomogeneity in Sec.~\ref{sec:model:inhom} and Sec.~\ref{sec:model:random}. Next, we proceed to the results of the numerical calculation of the dependence of the interaction energy on the correlation length. Our main results are summarized in Sec.~\ref{sec:summary}.

\section{Theoretical model}\label{sec:model}

\subsection{Homogeneous array}\label{sec:model:hom}
In this section, we present our approach to calculate the Casimir interaction  energy in a periodic array, mostly following Ref.~\cite{Bade_1957b}. The effects of array inhomogeneity and disorder will be analyzed in the following Sec.~\ref{sec:model:inhom} and Sec.~\ref{sec:model:random}. 

We consider a toy model of  $N$ {isotropic} molecules, periodically spaced with the distance $a$ between each other at the points $\bm R_n=na\bm{e}_z$, see Fig.~\ref{fig:schematics}. Each molecule is represented as a harmonic oscillator with a mass $m$, a displacement from the equilibrium position $\bm r_n=(x_n,y_n,z_n)$, an internal potential $kr_n^2/2$ with a spring constant $k$, and an {electric} dipole moment $pr_n$. 

Since our main goal is the effect of the disorder, we will include only the short-range electrostatic dipole-dipole interaction between the molecules and ignore the retarded long-range interaction.  The long-range interaction  can also be straightforwardly included into consideration by using the retarded electromagnetic Green function~\cite{Doniach_1963,Mahan_1965,Lucas1968,Renne1971}.
The coupling between the molecules is described by the following system of equations
\begin{equation}\label{eq:dipoles}
m\ddot{\bm r}_{n} + k\bm r_{n} + \frac{p^{2}}{3}
\sum\limits_{n'\ne n}\frac{3\bm R_{nn'} (\bm R_{nn'}\cdot\bm r_{n'})-\bm R_{nn'}^2\bm r_{n'}}{|\bm R_{nn'}|^5}=0\:,
\end{equation}
where $\bm  R_{nn'}=\bm R_n-\bm R_n'$.
In order to calculate the Casimir vibration energy, we look for the solutions of Eqs.~\ref{eq:dipoles}  with the harmonic time dependence, $\bm r_{n}(t)=
\bm r_{n}\e^{-\rmi\omega t}$ and obtain a linear system of equations for the $3N$ eigenfrequencies $\omega_\nu$.  The Casimir interaction energy is then found as an energy of zero-point vibrations with the subtracted energy of non-interacting molecules
\begin{equation}
    U=\frac{\hbar}{2}
\sum\limits_{\nu=1}^{3N}(\omega_\nu-\omega_0)\:,\label{eq:U0}
\end{equation}
 where $\omega_0=\sqrt{k/m}$. 
 {We note that the consideration of molecules as point dipoles is just an approximation. Corrections to the Casimir-Polder forces arising for the realistic extended molecules beyond this approximation are discussed in Ref.~\cite{Fiedler2015}.}
 
{Another notable approximation is that the considered model assumes that the molecules are suspended in a vacuum. In the more realistic system, such as a chain placed on a dielectric substrate, the electrostatic interaction at large distances will be screened~\cite{Glazov2018}. This is also termed as a shielding effect~\cite{Scheel2013}. However, the results we will focus on below will be mostly determined by the interaction between{ neighboring molecules, that are close to each other}, and the dielectric screening should not qualitatively affect  them}.

 A fully analytical solution even for such a simplified model for a finite $N$ is not possible because of the presence of the {long-range coupling}.  
An additional complication arises from the fact that all of the eigenmodes $\omega_\nu$ contribute to the interaction energy. Analytical calculation of all the eigenmodes in the general case of an inhomogeneous system does not seem feasible.
Since the dipole-dipole interaction quickly decays with distance, using a NN coupling approximation is reasonable. The summation in Eq.~\ref{eq:dipoles} in the NN approximation is restricted to the terms with $|n'-n|\le 1$. {The solution for $\bm r_n(t)$ can be sought as a simple standing wave $\bm r_n(t)=\bm A_k\sin(k n-\omega t)$ and the boundary conditions $\bm r_0=\bm  r_{N+1}=0$ yield $k=2\pi \nu/(N+1)$, with $\nu=1\ldots N$.}
The result for the Casimir interaction energy then assumes a closed form~\cite{Bade_1957b} 
\begin{multline}\label{eq:proto U}
U_{\rm{NN}}=\frac{E}{2}\sum_{\nu=1}^{N} \left[2\sqrt{1+2\beta\cos\left(\frac{\pi \nu}{N+1}\right)}\right.\\\left.+\sqrt{1-4\beta\cos\left(\frac{\pi \nu}{N+1}\right)}-3\:\right]\:,
\end{multline}
where we introduced the parameters
\begin{equation}\label{eq:E, beta}
E = \hbar \sqrt{\frac{k}{m}},\quad  \beta = \frac{p^{2}}{3ka^{3}}\:.
\end{equation}
The two terms in the square brackets of Eq.~\ref{eq:proto U} correspond to the vibrations in the transverse direction ($z_n=0$) and in the longitudinal direction ($\bm r_n\parallel z$), respectively. {Due to the  symmetry of the problem, the eigenmodes polarized along the chain direction $z$ are decoupled, and the modes polarized along $x$ and $y$ are degenerate with each other, resulting in the factor of $2$ before the first term in square brackets in Eq.~\ref{eq:proto U}}.

The dimensionless parameter $\beta$ characterizes the relative strength of the Casimir interaction as compared to the internal vibration energy.
If the interaction energy is assumed to be small, it is possible to further simplify Eq.~\ref{eq:proto U} by expanding it in the Taylor series in $\beta$~\cite{Bade_1957b}: 
\begin{equation}\label{eq:U}
U_{\rm{NN}}=-\frac{3E\beta^{2}(N-1)}{4}.  
\end{equation}
For $N=2$ this agrees with the standard expression for the interaction energy between the two particles~\cite{leonhardt2015forces}.
{We also note that for a periodic finite-length array, one can analytically evaluate not only the interaction energy but also the Casimir-Polder potential, describing the interaction of the array with an external object~\cite{Osestad2024}.}

Having recast these established results for a periodic chain, we now proceed to an inhomogeneous chain.
\subsection{Inhomogeneous array with the linear spatial variation of the coupling}\label{sec:model:inhom}

\begin{figure}[t]
\centering\includegraphics[width=\linewidth]{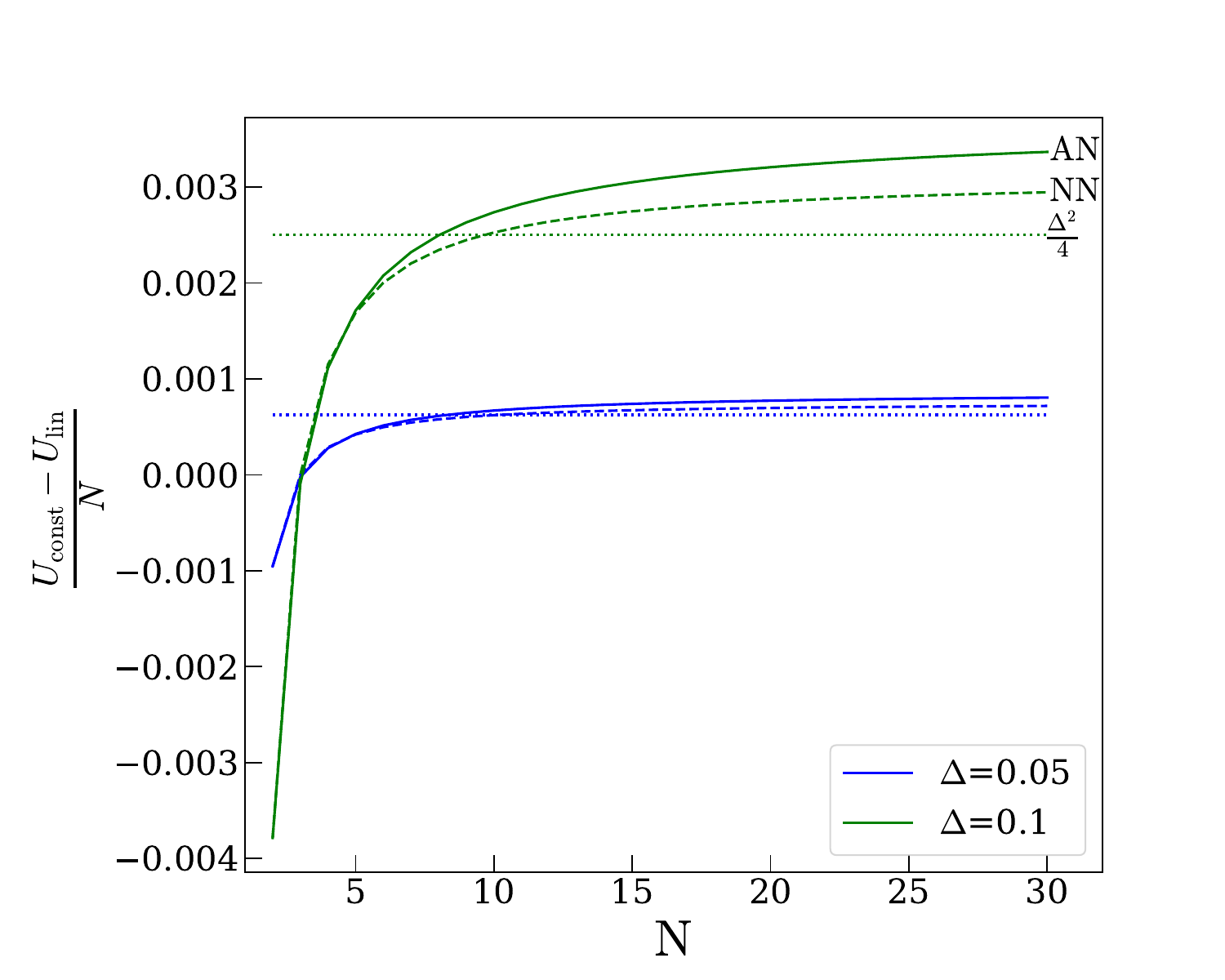}
\caption{Difference in energy per molecule between constant array $\beta(n) = \bar{\beta}$ values and array with linear distribution of $\beta(n)$
depending on the number of molecules ($N$). Calculation has been performed for  $\Delta=0.05$ (blue), $\Delta=0.1$ (green) for $E=1$ and $\bar{\beta}=0.1$. Solid lines present the results including the coupling between all neighbors (AN), dashed lines were calculated only with the nearest neighbor coupling (NN), and the dotted lines are the analytically derived asymptotes at $N\to \infty$ following Eq.~\ref{eq:Ulin}.} \label{fig:linear}
\end{figure}

We start from the most simple model, where the coupling parameter $\beta(n)$ is just a linear function of the coordinate $n$,
\begin{equation}\label{eq:linear}
\beta_{\rm lin}(n) = \bar{\beta}+\frac{2\Delta}{N-1}(n-1)  - \Delta,\quad n=1\ldots N\:,
\end{equation}
that is characterized by the mean value $\beta$ and the spread $\Delta$. 
{While this model is not quite realistic, it will provide useful insights for the model with random couplings $\beta(n)$, that will be considered in the next Sec.~\ref{sec:model:random}.

When the  dependence of the couplings Eq.~\ref{eq:linear} on the coordinate is linear,} it is in principle possible to solve the system of motion equations \ref{eq:dipoles} exactly, using the known answers for the Wannier-Stark ladder in solids~\cite{Mendez1993}. However, since we are interested only in the small values of $\beta$, it is easier to devise an even simpler approximation that builds on the perturbation theory result  Eq.~\ref{eq:U} for a homogeneous chain. We are interested in the interaction energy value per site $U/N$ for a large array. If the value of the spread of the interaction constants $\Delta$ is kept constant, for large $N$ the spatial variation of the coupling is slow. As such, we expect that for large $N$ one can use an adiabatic approximation and replace the factor $\beta^2$ in Eq.~\ref{eq:U} by its value {$\langle \beta^2\rangle \approx (\Delta^{2}/{3})+\bar{\beta}^{2}$}, averaged over the position in the chain. This results in 
\begin{equation}\label{eq:Ulin}
\langle {U}_{\rm{lin}}\rangle \approx -\frac{3E\bar{\beta}^{2}(N-1)}{4} - \frac{(N-1)\Delta^{2}}{4}\:.
\end{equation}

Figure~\ref{fig:linear} shows the interaction energy calculated depending on the size of the array $N$ for the two values of the spread of the interaction constants $\Delta\in\{0.05,0.1\}$. 
{The  calculation has been performed by substituting the harmonic time dependence $\bm r_{n}(t)=\bm r_{n}\e^{-\rmi\omega t}$ into 
Eqs.~\ref{eq:dipoles} generalized to include the spread of $\beta$.
We find the eigenfrequencies $\omega_\nu$ by numerical  diagonalization and evaluate the interaction energy from Eq.~\ref{eq:U0}.}
In order to elucidate the role of inhomogeneity, we subtract from the interaction energy its value $U_{\rm const}$ in the homogeneous array with $\beta(n)=\bar\beta$. The interaction energy per site monotonously grows with $N$ and saturates at larger $N$.
As expected, results from Eq.~\ref{eq:Ulin}, shown by dotted horizontal lines, provide a satisfactory approximation for the large $N$ limit in the nearest-neighbor approximation ({dashed} curves marked NN). It also reasonably well describes the results, including all neighbors ({solid} curves marked AN).
{The absolute difference between the AN and NN curves in Fig.~\ref{fig:linear} is more noticeable for a larger spread value $\Delta=0.1$ because the coupling value is larger. The relative difference of the AN and NN results for $N=30$ 
is about $10\%$ for $\Delta=0.05$ and $14\%$ for $\Delta=0.1$, while for $N=5$ the discrepancy is below $1\%$.  The higher discrepancy for $\Delta=0.1$ could be related to the $O(\Delta^{4})$ terms in the energy. Indeed, the number of terms in the perturbation expansion increases both with the perturbation order and with the number of couplings between the neighbors. As such, at higher $N$, the long-range coupling $O(\Delta^{4})$ terms can play a greater role.}

\subsection{Array with random coupling}\label{sec:model:random}

We now proceed to the array with the random couplings $\beta(n)$.  Before considering a large array, it is instructive to analyze the Casimir interaction energy in the particular case of $N=3$. In this case, considering the random variables $\beta(1), \beta(2), \beta(3)$ with a given mean value $\bar{\beta}$ and a given value of $E$, the calculation yields the following NN energy:
\begin{equation}\label{NN_N=3}
U_{\rm rand}^{N=3} = - \frac{3E}{4}\beta(2)[\beta(1)+\beta(3)]\:.
\end{equation}
For an array with a constant value of $\beta\equiv \bar{\beta}$ we would obtain 
$U_{\rm const}^{N=3} = {3E} - 3E\bar{\beta}^{2}/2$. This means that the energy modification due to the randomness of $\beta(n)$ is 
\begin{equation}
\langle U_{\rm const}^{N=3} - U_{\rm rand}^{N=3} \rangle = \frac{3E}{4} [\langle \beta(2)[\beta(1)+\beta(3)] \rangle- 2\bar{\beta}^{2}] \:.
\end{equation}

For uncorrelated disorder, in which the random variables $\beta(1)$, $\beta(2)$ and $\beta(3)$ are independent, we would expect $\langle U_{\rm const}^{N=3} - U_{\rm rand}^{N=3} \rangle = 0$. This means no difference in the interaction energy between the homogeneous array and the array with an uncorrelated disorder when $\langle \beta(n)\beta(n+1)\rangle=0$.
\begin{figure}[t!]
\centering\includegraphics[width=\linewidth]{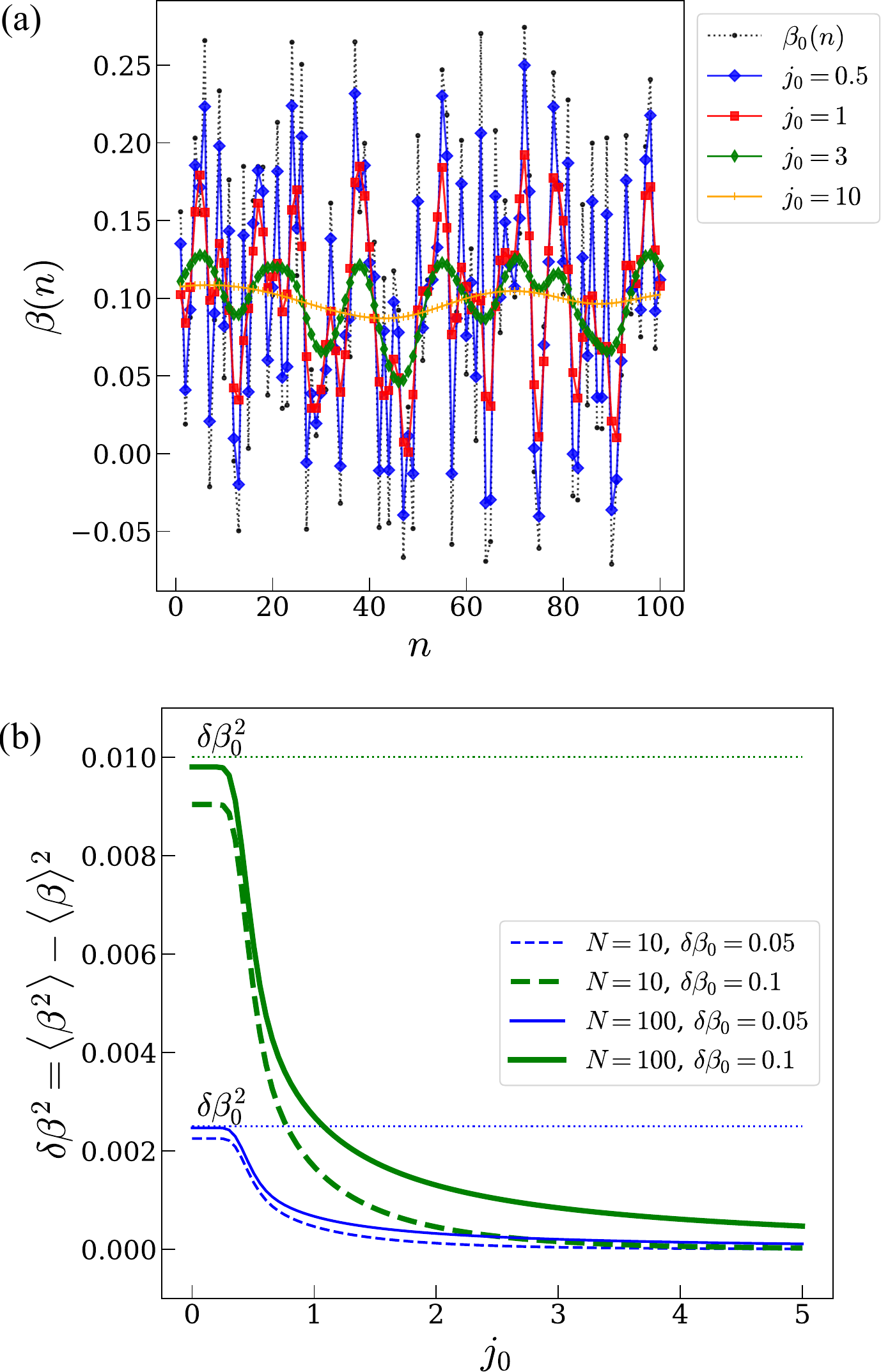}
\caption {(a) An example of the distribution of correlated $\beta(n)$ in different correlation lengths ($j_0$) as described in Eq.~\ref{eq:beta_rand} for ${\delta\beta_{0}=0.1}, \bar{\beta}=0.1, N=100$
{(b).} The dispersion of the correlated disorder distribution (averaged over 500 iterations) vs. the disorder correlation length $j_0$ for $N$=10, $N$=100 and {$\delta\beta_{0}=0.05$, $\delta\beta_{0}=0.1$}, when $\langle\beta_0\rangle=0.1$.
The dotted lines  {show the variance values for the uncorrelated uniform distribution $\beta_{0}$, which are by definition equal to $\delta\beta_{0}$}.
} \label{fig:examples}
\end{figure}

\begin{figure}[t]
\centering\includegraphics[width=\linewidth]{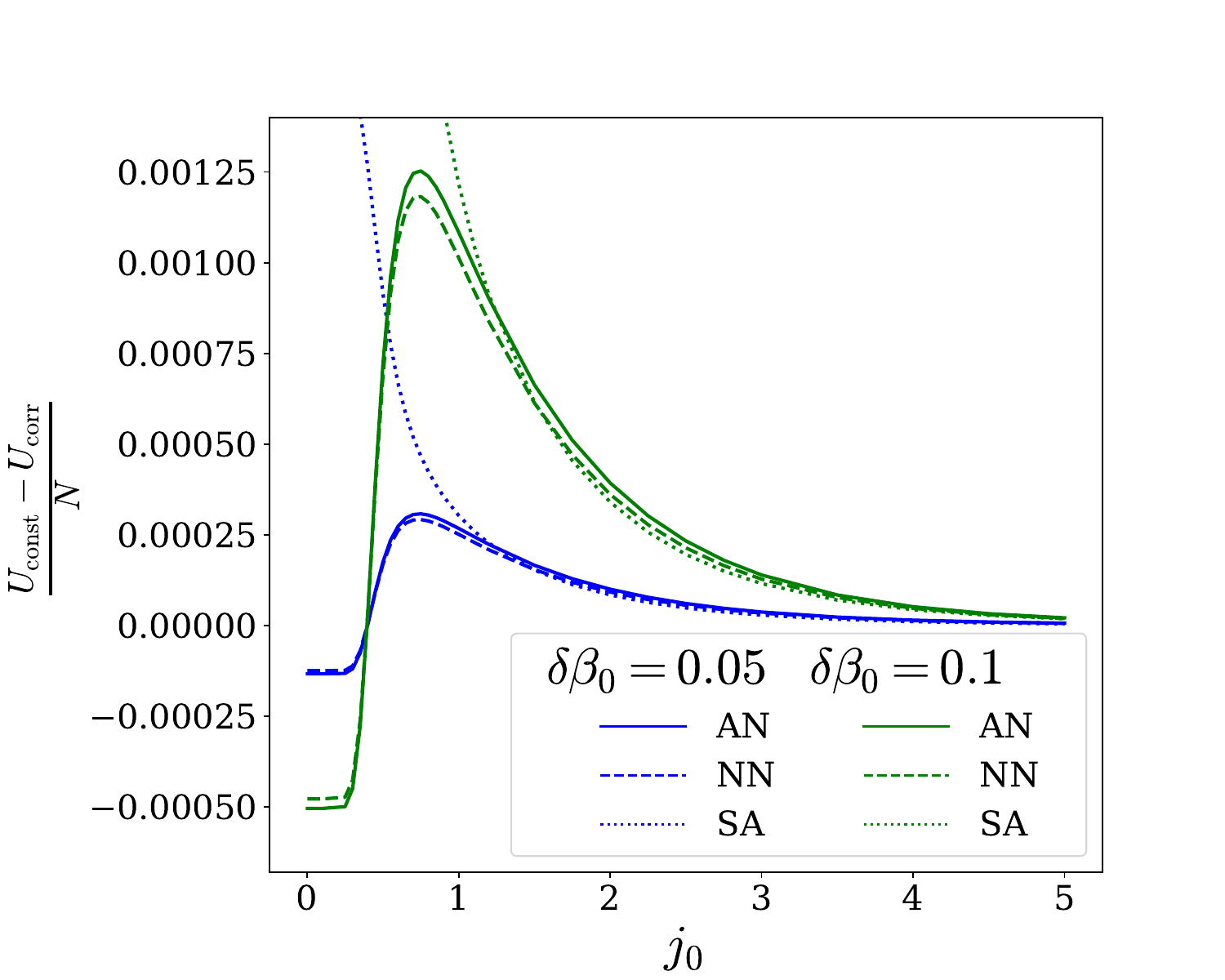}
\caption{ The difference in energy per molecule between the constant array with $\beta(n)\equiv\bar{\beta}$  and the array with correlated disorder depending on the correlation length $j_0$. Calculation has been performed for  $N=10$ molecules and {$\delta\beta_{0}=0.05$} (blue), {$\delta\beta_{0}=0.1$} (green) for $\bar{\beta}=0.1$ and $E=1$.  Solid lines are calculated including coupling between all neighbors, dashed lines are calculated with only nearest-neighbor coupling and dotted lines are semi-analytically estimated values based on calculated variance, see Eq.~\ref{eq: var. dependency}.} \label{fig:j0dep}
\end{figure}

 To analyze the role of the correlations in a more systematic way, we will now study a general random distribution of $\beta(n)$ with a finite correlation length $j_0$. We will define it as 
\begin{equation}\label{eq:beta_rand}
\beta(n)= \sum_{n'=1}^{N} \beta_0(n')\frac{e^{\frac{-(n-n')^2}{2j_0^2}}}{C}\:,
\end{equation}
where $\beta_0(n')$ are the independent  random variables 
uniformly  distributed in the range {$[\langle \beta_0\rangle-\sqrt{3}\delta\beta_0, \langle \beta_0\rangle+\sqrt{3}\delta\beta_0]$}, that is,  having the mean ${\langle\beta_0\rangle}$  and  the dispersion {$\delta\beta_0^2=\langle \beta_0^{2}\rangle-\langle \beta_0\rangle^2$}. The normalization coefficient $C$ is defined as
\begin{equation}\label{eq: normalization coefficient}
C = \sum_{n=-N}^{N} e^{-\frac{n^{2}}{2j_0^2}}\:.
\end{equation}
Examples of the random distribution $\beta(n)$ calculated following 
Eq.~\ref{eq:beta_rand} are shown in Fig.~\ref{fig:examples}(a). The calculation demonstrates that the distribution becomes smoother for larger $j_0$. At the same time, the span of the curves around their averaged value decreases. This results from the chosen normalization of the distribution Eq.~\ref{eq: normalization coefficient}. The dispersion decrease is  confirmed by the results Fig.~\ref{fig:examples}(b). This panel presents the numerically calculated dispersion of the correlated disordered array
$\delta\beta^2=\langle \beta^2\rangle-\langle \beta\rangle^2$ depending on the disorder correlation length $j_0$ for two different values of ${\delta\beta_{0}}$. We also show the results for two different values of $N=10$ and $N=100$. As expected, if the disorder correlation length is much smaller than unity, the values of $\beta(n)$ for different values of $n$ are not correlated with each other, so that the dispersion does not depend on the correlation length and is equal to that of the uncorrelated distribution (dotted lines).

Next, we show in Fig.~\ref{fig:j0dep} our main result: the dependence of the Casimir interaction energy on $j_0$. Just like in Fig.~\ref{fig:linear}, we subtract from the interaction energy its value in the homogeneous array with $\beta(n)=\bar\beta.$  The  calculations are performed for the two different values of {$\delta\beta_{0}$}. {The dependence of the interaction energy on {$\delta\beta_{0}$} is an approximately quadratic one: the green curves, where {$\delta\beta_{0}=0.1$}, have approximately four times larger  span than the blue curves with the twice lower value of {$\delta\beta_{0}=0.05$.}}

The most important result in Fig.~\ref{fig:j0dep} is the nonmonotonous dependence of the interaction energy on $j_0$ with a peak around the value $j_0=1$. This behavior can be explained in the following way. For small value of $j_0$ the disorder is uncorrelated, $\langle\delta\beta(n)\delta\beta(n+1)\rangle=0$. As such, based on our analysis of the small cluster with $N=3$ above, the {$\delta\beta_{0}^2$}-terms in the Casimir interaction energy are suppressed. 
Similarly to Fig.~\ref{fig:linear}, we show in Fig.~\ref{fig:j0dep} two sets of numerical results, calculated including the coupling between all neighbors and only nearest-neighbor coupling. However, these results are quite close, so the energy mostly depends on the short-range coupling. The observed suppression of the coupling energy for small $j_0$ is consistent with our analysis of the small cluster with $N=3$ above.
We note, however, that the interaction energy is still nonzero for $j_0\to 0$. We can explain this {by either the terms of higher order in $\beta^2$}, beyond the approximation used in Eq.~\ref{NN_N=3}, or by the presence of long-range couplings.

In the opposite limiting case of smooth disorder, $j_0\gg 1$,  dispersion ${\delta\beta^{2} =} \langle \beta^2\rangle-\langle \beta\rangle^2$ decreases with $j_0$, see also Fig.~\ref{fig:examples}(b). As such, we can expect a decrease in the difference in the interaction energy from its value in the homogeneous array. Moreover, it is also possible to obtain a semi-analytical approximation for the interaction energy in the case when $j_0\gtrsim 1$. {To this end, we 
use the results in the previous Sec.~\ref{sec:model:inhom} obtained for the linear distribution of $\beta(n)$. Namely, we replace the results for the random model with given values of $\langle\beta\rangle$ and {$\delta\beta$} by the results obtained for the linear distribution $\beta(n)$ with the same mean $\langle\beta\rangle\equiv \bar\beta$ and the same second moment 
$\langle\beta^2\rangle$. This means that the dispersion $\delta\beta^2$ in the random model is related to the spread in the linear model as $\delta\beta^2=\Delta^2/3$.} In this approximation, we
replace in Eq.~\ref{eq:Ulin} the value of $\Delta^2/3$ by the dispersion $\langle \beta^2\rangle-\langle \beta\rangle^2$ in the array with a finite correlation length and obtain for $N\gg 1$
\begin{equation}\label{eq: var. dependency}
 \frac{\langle U_{\rm corr}\rangle}{N}  = -\frac{3E\bar\beta^{2}}{4} -\frac{3E}{4} ( \langle \beta^{2} \rangle -  \bar{\beta}^{2})\:.
 \end{equation}  
The corresponding results  are shown in Fig.~\ref{fig:j0dep} by thin dotted curves and well describe the results of the full numerical calculation for $j_0\gtrsim 1$.

\section{Conclusions}\label{sec:summary}
To conclude, we have conducted a theoretical study of van der Waals interaction forces in one-dimensional chains described by a coupled dipole model with a correlated disorder. We have focused on the dependence of the coupling energy on the disorder correlation length $j_0$. The dependence is nonmonotonous and has a single maximum at  $j_0\approx 1$. For larger values of $j_0$, the interaction energy is suppressed and can be described by an analytical approximation obtained within the limit of smooth disorder. For smaller values of $j_0$, the interaction energy is also suppressed in agreement with our simplified analysis of a small cluster with $N=3$ dipoles.

Our results indicate that the problem of van der Waals interactions is not trivial even for the simplest one-dimensional setting. A natural extension of this study would be to consider higher dimensional disordered models and to include the effects of retardation.

\acknowledgements
We are grateful to Ulf Leonhardt for useful discussions. ANP acknowledges support by the Center for New Scientists at the Weizmann Institute of Science.  YM has been supported by the ``Young Scholars'' program of the Weizmann Institute of Science.


%

\end{document}